\documentclass[12pt,a4paper,leqno]{article}
\usepackage{graphicx}
\usepackage[OT2,OT1]{fontenc}
\DeclareTextSymbol{\i}{OT2}{26}
\DeclareTextSymbol{\I}{OT2}{18}
\newcommand{\cyr}{\fontencoding{OT2}\selectfont}
\newcommand{\lat}{\fontencoding{OT1}\selectfont}
\title{\sc On correlations and fractal characteristics of time series }
\author{Nikolay K. Vitanov$^{1}\footnote{coresponding author}$,  Kenshi Sakai$^{2}$ and
Elka D. Yankulova$^{3}$}
\date{${1}$ Institute of Mechanics, Bulgarian Academy of Sciences,
Akad. G. Bonchev Str., Bl. 4, 1113, Sofia, Bulgaria, e-mail:
vitanov@imbm.bas.bg \\
$^{2}$ Tokyo University of Agriculture and Technology, 3-5-8, Saiwai-cho,
Fuchu-shi, Tokyo, 183-8509, Japan, e-mail: ken@cc.tuat.ac.jp \\
$^{3}$ Faculty of Biology, "St. Kliment Okhridski" University of Sofia,
8, Dragan Tsankov Blvd., 1162 Sofia, Bulgaria, e-mail: eyankulova@yahoo.com}
\begin{document}
\pagenumbering{roman}
\maketitle
\begin{abstract}
Correlation analysis is convenient and frequently used tool for
investigation of time series from complex systems. Recently new
methods such as the multifractal detrended fluctuation analysis
(MFDFA) and the wavelet transform modulus maximum method (WTMM)
have been developed. By means of these methods (i) we can
investigate long-range correlations in time series and (ii) we can
calculate  fractal spectra of these time series. But opposite to
the classical tool for correlation analysis - the autocorrelation
function, the newly developed tools are not applicable to all
kinds of time series. The unappropriate application of MFDFA or
WTMM leads to wrong results and conclusions. In this article we
discuss the opportunities and risks connected to the application
of the MFDFA method to time series from a random number generator
and to experimentally measured time series (i) for accelerations
of an agricultural tractor and (ii) for the heartbeat activity of
{\sl Drosophila melanogaster}. Our main goal is to emphasize on
what can be done and what can not be done by the MFDFA as tool for
investigation of time series.
\end{abstract}
Key words: nonlinear time series analysis, fractals, Hurst exponent,
H{\"o}lder exponent
\newpage
\begin{abstract}
\cyr
Korelacionniyat analiz e udoben i qesto izpolzvan instrument
za izsledvane na vremevi redove ot sloz1ni sistemi. V poslednite
godini byaha razviti novi metodi za analiz na korelacii kato
naprimer multifraktalniyat  fluktuacionen analiz
(\lat MFDFA) \cyr ili metodp2t na maksimumite na modulite na
\lat wavelet -\cyr transformaciyata (\lat WTMM). \cyr Qrez tezi
metodi \lat (i) \cyr nie moz1em da izsledvame korelaciite s dp2lp2g
obseg vp2v vremevi redove i \lat (ii) \cyr da izqislyavame fraktalnite
spektri na veliqinite, harakterizirawi tezi vremevi redove. No
protivno na sluqaya s klasiqeskiya instrument za korelacionen analiz -
avtokorelacionnata funkciya, gorespomenatite metodi ne sa priloz1imi za
vsyakakvi vremevi redove, a bezrazbornoto im prilagane moz1e da dovede
do grexni rezultati i zaklyuqeniya. V tazi statiya nie obsp2z1dame
vp2zmoz1nostite i riskovete, svp2rzani s prilaganeto na \lat MFDFA
\cyr metoda kp2m vremevi redove ot edin generator na sluqa\i ni
qisla i kp2m eksperimentalno izmereni vremevi redove ot \lat (i) 
\cyr uskoreniya na traktor i \lat (ii) \cyr ot sp2rdeqnata de\i nost
na \lat {\sl Drosophila melanogaster}. \cyr Naxata glavna cel e da
podqertaem kakvo moz1e i kakvo ne moz1e da se napravi s \lat MFDFA
\cyr metoda kato instrument za izsledvane na vremevi redove.   
\lat
\end{abstract}
\newpage
\pagenumbering{arabic}
\section{Introduction}
The investigation of correlation properties is one of the first
tasks that a researcher of a time series has to perform.  The
classical tool for this is the autocorrelation function which can
be calculated for stationary as well as for the nonstationary time
series. The autocorrelation function reflects the behaviour of the
time series. If for an example a time series is periodic its
autocorrelation function is periodic too. For the most
experimentally measured time series however the correlation
function decays and in simple systems this decay in the most of
the cases is exponential one. Opposite to this in numerous complex
systems, which consist of many interacting subsystems, the
correlations decay with power law and are
called long-range because at large time scales the power law
function is always large than the exponential function.
\par
The properties of the long-range correlations can be investigated by
means of two recently developed methods: WTMM and MFDFA. WTMM is based
on the properties of the wavelets \cite{muzy91,muzy93,phi99,dv} and it leads to
excellent results when the recorded time series are long enough.
For shorter time series (usually below $10^{6}$ values) more convenient
is the much simpler for implementation MFDFA method based on the scaling
properties of the so called fluctuation function.
Below we shall discuss what
can be done and what can not be done by the MFDFA method when it is
applied to different kinds of time series. We use two kinds of experimentally
recorded time series (i) three short time series for the acceleration of
an agricultural tractor working in different regimes and (ii) two time
series for the heartbeat activity of {\sl Drosophila melanogaster}.
In addition we shall investigate a time series obtained by a random
numbers generator in order to illustrate the monofractal properties
of a time series.
\par
The paper is organized as follows. In the next section we discuss
some multifractal quantities and how they can be calculated by the
MFDFA method. In section 3 we  give more information about the
time series and analyze their autocorrelation functions. In
section 4 we analyze the time series by means of the MFDFA method.
Several concluding remarks are summarized in the last section.
\section{Generalized dimensions, fractal spectra, H{\"o}lder and Hurst
exponents} 
Let us have a set S which lies in an $N-$dimensional
Cartesian space covered by a grid of $N$-dimensional boxes of edge
length $\epsilon$. Let $\epsilon$ be small and we need
$N^{*}(\epsilon)$ boxes to cover our set. The box-counting
dimension of the set is
\begin{equation}\label{bcountd}
D_{0}=\lim_{\epsilon \to 0} \frac{\ln N^{*}(\epsilon)}{\ln (1/\epsilon)}
\end{equation}
$D_{0}$ is not an integer for some sets. These sets are called fractals.
$D_{0}$ is a member of the spectrum of the $D_{q}$ dimensions
\begin{equation}\label{dqdim}
D_{q}=\frac{1}{1-q} \lim_{\epsilon \to 0} \frac{\ln I(q, \epsilon)}{
\ln (1/\epsilon)}
\end{equation}
where $q$ is a continuous index and $I(q,\epsilon)$
\begin{equation}
I(q,\epsilon) = \sum_{k=1}^{N^{*}(\epsilon)} \mu_{k}^{q},
\end{equation}
is a sum over the $N^{*}$ boxes of size $\epsilon$ which cover the
set. $\mu_{k}$ is the natural measure i.e. it is a measure of the
frequency with which typical orbit visit various boxes covering
the attractor for the limit case when the length of the orbit goes
to infinity (in addition the frequencies have to be the same for
all initial conditions in the basin of attraction of the attractor
except for a set with Lebesque measure $0$). Thus for $\mu_{k}$ we
have
\begin{equation}\label{muk}
\mu_{k}=\lim_{T \to \infty} \frac{\xi (c_{k},{\bf x}_{0},T)}{T},
\end{equation}
where $\xi$ is the time the orbit originating from ${\bf x}_{0}$
spends in the cube $c_{k}$ in the time interval $0 \le t \le T$.
From (\ref{dqdim}) by means of the L`Hospital rule we easy can
obtain
\begin{equation}\label{d1}
D_{1}=\lim_{\epsilon \to 0} \frac{\sum_{k=1}^{N^{*}}(\epsilon) \mu_{i}
\ln \mu_{i}}{\ln \epsilon}
\end{equation}
$D_{1}$ is called also information dimension. In general $D_{0}
\ge D_{1} \ge D_{2} \ge \dots$. If $D_{q}$  varies with $q$ the
measure, associated with $D_{q}$ is called multifractal measure.
The measure can have different behavior in each cube covering the
set regardless of how small is its size $\epsilon$. Thus the
dimension is not enough for a characterization of the measure and
we need additional characteristic quantities. One such quantity is
the coarse-grained H{\"o}lder exponent for each box
\begin{equation}\label{hold1}
\alpha=\frac{\ln \mu(box)}{\ln \epsilon}
\end{equation}
$\mu$ can have irregular behavior for different boxes. When the
box size decrease and $\mu$ has statistically the same irregular
behavior the measure is called self-similar \cite{peitgen92}. For
large class of such measures   $\alpha$ lies
between $\alpha_{min}>0$ and $\alpha_{max}<\infty$. For given
$\epsilon$ we can count the number $N_{\epsilon}$ of the boxes
with value of $\alpha$ in small interval around a prescribed
value. Thus we can define the coarse-grained spectrum
\begin{equation}\label{cgspectrum}
f_{\epsilon}(\alpha) \propto - \frac{\ln N_{\epsilon}(\alpha)}{\ln \epsilon}
\end{equation}
From (\ref{cgspectrum}) we obtain $N_{\epsilon}(\alpha) \approx
\epsilon^{-f(\alpha)}$, i.e. the relationship which has a form
analogous to this one for the dimension above. Thus $f(\alpha)$
can be treated as fractal dimension of the subsets $S_{\alpha}$ of
$S$ having coarse grained H{\"o}lder exponent equal to $\alpha$.
If $f(\alpha)$ is not constant then for small $\epsilon$ the set
$S$ is constructed by subsets $S_{\alpha}$ of different dimensions
$f(\alpha)$. Then the set $S$ is called multifractal.
\par
The multifractals can be characterized by different spectra. Below
we shall obtain relationships for these spectra for the case
$\epsilon \to 0$. We remember that the set $S$ be covered by grid
of boxes of unit size $\epsilon$ and $\mu$ is the probability
measure on $S$ ($\mu(S)=1$). Let $\mu(c_{k})=\mu_{k}$ where
$c_{k}$ denotes again the $k-$th cube. We have assigned a
singularity measure $\alpha_{k}$ to each cube
\begin{equation}\label{singmeas}
\mu_{k}= \epsilon^{\alpha_{k}}
\end{equation}
For small $\epsilon$ we can make continuous approximation for the
number of cubes for which $\alpha_{k}$ is between $\alpha$ and $\alpha+
d \alpha$, i.e., we can denote this number as
\begin{equation}\label{contappr}
\rho(\alpha) \epsilon^{-f(\alpha)} d \alpha
\end{equation}
By substitution of (\ref{singmeas}) in the relationship for $I(q,\epsilon)$
and after a transition from a sum over the boxes to an integration over
the $\alpha$ we obtain
\begin{eqnarray}\label{i1}
I(q,\epsilon)= \sum_{k=1}^{N^{*}(\epsilon)} \epsilon^{\alpha_{k} q} =
\int d \alpha^{*} \rho (\alpha^{*}) \epsilon^{-f(\alpha^{*})}
\epsilon^{q \alpha^{*}} =\nonumber \\
=\int d \alpha^{*} \rho (\alpha^{*}) \exp \left \{ [f(\alpha^{*})-
q \alpha^{*}] \ln (1/\epsilon) \right \}
\end{eqnarray}
For small $\epsilon$ $\ln (1/\epsilon)$ is large and the main
contribution to the above integral is from the neighborhood of the
maximum value of the $f(\alpha^{*})-q \alpha^{*}$. Let
$f(\alpha^{*})$ be smooth and the maximum is located at
$\alpha^{*}=\alpha(q)$ given by
\begin{equation}\label{max1}
\frac{d}{d \alpha^{*}} [f(\alpha^{*})-q \alpha^{*}] \mid_{\alpha^{*}=
\alpha(q)}=0 \to \frac{d f}{d \alpha^{*}}\mid_{\alpha^{*}=\alpha}=q
\end{equation}

\begin{equation}\label{max2}
\frac{d^{2}}{d (\alpha^{*})^{2}} [f(\alpha^{*})-q \alpha^{*}] \mid_{\alpha^{*}=
\alpha(q)}=0 \to \frac{d^{2} f}{d (\alpha^{*})^{2}}\mid_{\alpha^{*}=\alpha}=q
\end{equation}
Now we take the Taylor series representation of the
function $F(\alpha^{*},q)=f(\alpha^{*})-q\alpha^{*}$ around the point
$\alpha^{*}=\alpha (q)$ and  substitute it in (\ref{i1}). The result is
\begin{eqnarray}\label{i2}
I(q,\epsilon)=\exp \left\{ \right [f(\alpha(q))-q \alpha] \ln(1/\epsilon)\}
\int d \alpha^{*} \rho (\alpha^{*}) \epsilon^{-(1/2)f^{ \prime \prime}
(\alpha(q)) (\alpha^{*}-\alpha(q))^{2}} \nonumber \\
\approx \exp \left\{ \right [f(\alpha(q))-q \alpha] \ln(1/\epsilon)\}
\end{eqnarray}
Introducing (\ref{i2}) in (\ref{dqdim}) we obtain
\begin{equation}\label{dqdim2}
D_{q}=\frac{1}{q-1} \left[ q \alpha(q) - f(\alpha(q)) \right]
\end{equation}
From (\ref{max1})
\begin{equation}\label{dqx}
\frac{d}{d q} \left[ (q-1) D_{q} \right] = \alpha (q) = \frac{d \tau}{d q}
\end{equation}
Then
\begin{equation}\label{tauq}
\tau(q) = (q-1) D_{q} \to D_{q}= \frac{\tau (q)}{q-1}
\end{equation}
From (\ref{dqdim2})
\begin{equation}\label{f2}
f(\alpha(q)) = q \frac{d \tau}{d q} -(q-1) D_{q} = q \frac{d \tau}{d q} -
\tau (q)
\end{equation}
Thus for each $q$ from (\ref{tauq}) and (\ref{f2}) give $\alpha
(q)$ and $f(\alpha)$ thus parametrically specifying the function
$f(\alpha)$. The Hurst exponent sometimes is associated with the
coarse-grained H{\"o}lder exponent for $\epsilon \to 0$. This can
be done in the following way. For the cases in which the following
relationship holds
\begin{equation}\label{th}
\tau(q) = q h(q) -1
\end{equation}
we obtain
\begin{equation}\label{a2}
\alpha(q) = \frac{d \tau}{dq}= h(q) + q \frac{d h}{dq}
\end{equation}
and
\begin{equation}\label{falpha}
f(\alpha)=q \alpha - \tau(q) = q [\alpha - h(q)] +1
\end{equation}
 But in which cases holds (\ref{th})? Let us consider stationary
 time series $\{x_{k}\}, k=1, \dots, N$ and let us use appropriate
transformations in order to make all values positive and to
normalize the time series, i.e.,
\begin{equation}\label{normaliz}
x_{k} \ge 0, \hskip.5cm \sum_{k=1}^{N} x_{k}=1
\end{equation}
In this case we can associate the time series with some probabilities
and we shall use this fact to derive (\ref{th}). Let us construct the
profile function for our normalized time series
\begin{equation}\label{prof1}
Y_{n}=\sum_{k=1}^{n} (x_{k} - \langle x \rangle)
\end{equation}
where $\langle x \rangle$ is the mean of the time series. Now we divide
the time series into $N_{s}$ segments of length $s$ and let for
simplicity $N_{s}=N/s$ is an integer. The sum
\begin{equation}\label{sum1}
Y(\nu s) - Y((\nu - 1) s) = \sum_{k=(\nu-1)s+1}^{\nu s}
(x_{k} - \langle x \rangle )
\end{equation}
for the segment $\nu$ is identical to the box probability $p_{s}(\nu)$
which is the main building block of the corresponding partition sum
$Z_{q}(s)$ and scales with the $\tau_{q}$ spectrum
\begin{equation}\label{scaling1}
Z_{q}(s) = \sum_{\nu=1}^{N/s} \mid p_{s}(\nu) \mid^{q} \propto
s^{\tau(q)}
\end{equation}
From other side for our time series $\{x_{k} \}$ we can build a
scaling function which contains the local Hurst exponent. This is the
sum
\begin{equation}\label{sum2}
\left \{  \frac{1}{2 N_{s}} \sum_{\nu=1}^{2N_{s}} \mid Y(\nu s) - Y((\nu-1) s
\mid^{q}\right \}^{1/q} \propto s^{h(q)}
\end{equation}
From here we obtain
\begin{equation}\label{relat1}
  \frac{s}{2 N_{s}} \sum_{\nu=1}^{2N/s} \mid Y(\nu s) - Y((\nu-1) s
\mid^{q} \propto s^{qh(q)}
\end{equation}
and finally
\begin{equation}\label{relat2}
   \sum_{\nu=1}^{N/s} \mid Y(\nu s) - Y((\nu-1) s
\mid^{q} \propto s^{qh(q) -1}
\end{equation}
A comparison of (\ref{scaling1}) and (\ref{relat2}) leads to
(\ref{th}).
\par
The implementation of the  MFDFA method  \cite{kant} follows the steps below.
The starting point are our time series $\{x_{k}\}$ of finite length $N$. First of all we
have to determine the profile function. Here we have to mention the following. There
are two possibilities. First of all we can use profile function as
in (\ref{prof1}), i.e.,
\begin{equation}\label{profile1}
Y(i) = \sum_{k=1}^{i} (x_{k} - \langle x \rangle)
\end{equation}
The final result of our analysis will be to obtain some values of the
local Hurst exponent $h$. If we use (\ref{profile1}) as profile
function we can determine only positive Hurst exponents which are not
quite close to $0$. In all other cases we have to use as profile
\begin{equation}\label{profile2}
\hat{Y}(i) = \sum_{k=1}^{i} (Y(k)- \langle Y \rangle)
\end{equation}
where $Y(k)$ comes from (\ref{profile1}). Thus we shall obtain
value of local Hurst exponent which is larger than the value of
true exponent i.e.
\begin{equation}\label{hgen}
\hat{h}(q)=1+h(q)
\end{equation}
This is the final difference between the two branches of the
method. We shall present here the first branch, which has the same
steps as the second with the only difference that we have
everywhere to change $Y$ with $\hat{Y}$ if we want to obtain the
algorithm for the second branch. After the calculations of the
profile function we have to divide it into segments of length $s$.
As $N/s$ in general is not an integer some part at the end of the
time series will remain out of the $N_{s}=int(N/s)$ segments. In
order to incorporate the influence of this part into the analysis
we divide the time series from the end to the beginning again to
$N_{s}$ segments of length $s$. Thus we obtain $2N_{s}$ segments
and for each segment calculate the local polynomial trend
$y_{\nu}(p)$, $p=1,\dots,s$. The trend can be a polynomial of
order $m$. Thus we perform MFDFA of order $m$ and the coefficients
of the polynom are obtained by a least-square fit of the
corresponding segment. After the trend calculation we determine
the variance
\begin{equation}\label{variance1}
F^{2}(\nu,s) = \frac{1}{s} \sum_{i=1}^{s} \{Y[(\nu-1)s+i]-y_{\nu}(i) \}^{2}
\end{equation}
for the segments $\nu=1,2,\dots,N_{s}$. For the segments $\nu=N_{s}+1,N_{s}+2,\dots,2 N_{s}$
the variance is
\begin{equation}\label{variance2}
F^{2}(\nu,s)= \frac{1}{s} \sum_{i=1}^{s}  \{Y[N-(\nu-N_{s})s+i] - y_{\nu}(i) \}^{2}
\end{equation}
The next step is an averaging over the segments for obtaining the $q$-th order fluctuation function
($q$ is in general a real number). For $q=0$ the fluctuation function is
\begin{equation}\label{zerothord}
F_{0}(s)=\exp \left \{ \frac{1}{4N_{s}} \sum_{\nu=1}^{2N_{s}} \ln [ F^{2}(\nu,s)] \right \}
\end{equation}
and for $q \ne 0$
\begin{equation}\label{nonzero}
F_{q}(s) = \left \{ \frac{1}{2N_{s}} \sum_{\nu=1}^{2 N_{s}} [F^{2}(\nu,s)]^{q/2} \right \}^{1/q}
\end{equation}
For large class of time series the fluctuation function has a
power law behavior and from this power law for large $s$ we can
determine the local Hurst exponent
\begin{equation}\label{scal1}
F_{q}(s) \propto s^{h(q)}
\end{equation}
\section{The time series and their autocorrelation functions}
In order to achieve our goals we have selected time series from
three different systems: an agricultural tractor, time series from
the heartbeat activity of the classical object of Genetics -{\sl
Drosophila melanosgaster} and a time series from a random number
generator in order to illustrate the case of absence of
correlations as well as the  monofractal properties of some time
series. The investigated time series are shown in Fig. 1. We
investigate two kinds of experimental time series for acceleration
of agricultural tractors. The
time series from panels (a) and (b) of Fig. 1  are from
experimental modeling of the bouncing phenomenon. It arises when a
machine travels on a rough road and consists of large-amplitude
oscillations which can lead to injuries of the driver, to a lift
of the machine (some times more than half-a-meter over the
surface) and to damage of some of its elements. The time series
for this phenomenon are recorded when the tractor moved on
artificial rough road, consisting of 12 small kinks prepared on am
asphalted surface. For this case we have several large amplitude
oscillations when the tractor moves on the rough part of the road
and in some time series we have also the noise background when the
tractor moves on the smooth part of the road. Thus we can discuss
these time series as time series of noise with superimposed
periodic trend. The time series from panel (c) of Fig. 1
are recorded when the tractor has a construction exhibiting impact
properties. This case has been chosen because the impact systems
are frequent source of chaotic vibrations \cite{ken94},
\cite{ken99}. The tractor has been converted to an impact system
by adding vibrating subsoiler to it. The soil cutting chiesel
breaks a hard soil layer located at the depth between 30 and 50 cm
from the field surface and it is oscillated in order to reduce the
draft force and to improve the water infiltration into the soil.
The oscillation is realized by a hydravlic power cylinder with two
lift arms. This work mode is called ground-penetrating mode. 
\par
The time series in panels (d) and (e) of Fig. 1 are records of the
heartbeat activity of {\sl Drosophila melanogaster} These flies
are provided by Bloomington Drosophila Stock Center, U.S.A. The time
series are ECGs (electrocardiogramms) for the first generation
(the kids) obtained by male Dopa decarboxilase
(Ddc) mutant (FBgn 0000422) located in chromosome 2, locus 37C1,
crossed with female shibire (shi)  (FBgn 0003392) located in chromosome
1, locus 13F7-12. Ddc codes for an enzyme necessary for the
synthesis of four neurotransmitters: norepinephrine, dopamine,
octopamine, serotonin, related to learning and memory. The shibire
(shi) mutants cause paralysis at high temperature and  eliminates the effect
of the neurotransmitters on the heart. The ECGs have been recorded optically
at a stage P1 (white puparium) of a Drosophila
development when it is both immobile and transparent and the
dorsal vessel is easily viewed. The object was placed on a glass
slide in a drop of distilled water under a microscope
(magnification 350 x). Fluctuation in light intensity due to
movement of the dorsal vessel tissue was captured by photocells
fitted to the one eyepiece of the microscope. The captured
analogue signal was then digitized at 1 kHz sampling rate by data
acquisition card and LabVIEW data capturing software supplied by
National Instruments. Finally the time series from the random number
generator are obtained by means of the generator {\sf ran2} from
\cite{nrec}.
In Fig. 2 are presented the autocorrelation functions for the
investigated time series. The large amplitude oscillations in
panels (a) and (b) reflect the bouncing phenomenon. The impact
oscillations are clearly reflected by the almost periodic
behaviour of the autocorrelation function in panel (c). The slowly
decaying autocorrelations for the ECGs of {\sl Drosophila} show
some degree of anticorrelation at large $n$. Finally as it can be
expected the time series from the random number generator are
uncorrelated.
\section{Fractal analysis of the time series}
Excluding the time series from the random number generator which
are uncorrelated, the other time series exhibit some degree of
long range correlations. Below we shall discuss the question how
far these correlations can be investigated by means of the modern
methods such as the MFDFA.
In order to answer this question we have to look at the
fluctuation functions connected to the time series. We shall
calculate the fluctuation functions directly on the basis of the
time series for the cases of tractor accelerations and for the
time series from the random number generator. For the time series
from Drosophila we first shall construct the time series for the
intermaxima intervals (the time series for the time between two
consequent maxima of the time series) and for the intermaxima time
series we shall calculate the fluctuation functions and eventually
the characteristic fractal quantities. The results are shown in
Fig. 3. For the  time series for the acceleration of the tractor
in bouncing regime we can not apply the MFDFA method. The reason
is that these time series (shown in panels (a) and (b) in Fig. 1)
are too short and we can not divide them in large number of
segments in order to have enough statistics for correct
calculation of the fluctuation function. In order to apply the
MFDFA the time series must have at least two thousand values. The
minimum number of values - 4000 for which the fluctuation function
is calculated here are the time series for the acceleration of
the tractor in impact regime. The fluctuation function is not a
straight line on a log-log scale - it has a kink as a reflection
of periodicity of the time series. Thus these time series do not
show scaling behaviour and we can not calculate the fractal
spectra for them. On the other side we confirm the observation
\cite{hu2001} that the periodic behaviour influences the
fluctuation function. On panels (b) and (c) of Fig. 3 are
presented the fluctuation functions for the intermaxima intervals
of Drosophila. On panel (b) the fluctuation function is close to a
straight line, i.e., the scaling properties can be assumed and the
multifractal spectra can be calculated. Not so clear is the
situation in panel (c). We shall carry further the calculations
for this case in order to see how the not very large deviations
from the scaling behaviour influence the form of the fractal
spectra. The best scaling properties are exhibited by the time
series from the random number generator - panel (d). This can be
expected as these time series are specially selected for
illustration of monofractal behaviour.
Thus for calculation of the fractal characteristics from the
initial 6 time series  the requirements for length and presence of
scaling properties have eliminated 3 time series. The $h(q)$
spectra for the remaining three time series are presented in Fig. 4. The
expected behaviour of such kind of spectrum is presented in panel
(a) where $h$ decreases with increasing $q$. This multifractal
behaviour reflects the scaling properties of the time series as
shown in panel (b) of Fig. 3. Multifractal behaviour is exhibited
also by the second time series for the intermaxima intervals of
{\sl Drosophila melanogaster} shown in panel (b) of Fig. 4. As it
can be seen from panel (c) of Fig. 4  for the time series from the
random number generator $h$ is a constant for all $q$ which is
characteristic feature for the monofractal behaviour.
\par
Finally we calculate the $f(\alpha)$ spectra. For the time series
from the random number generator this spectrum is a point and it
is not presented in Fig. 5. Panel (a) shows the spectrum for the
time series with scaling properties shown in panel (b) of Fig. 3. 
The expected form for the
$f(\alpha)$ spectrum is parabolic one. This form is observed on
panel (a) of Fig. 5. The maximum of the $f(\alpha)$ spectrum shows
at which $\alpha$ is positioned the most statistically significant
part of the time series, i.e., the subsets with maximum fractal
dimension among all subsets of the time series. In our case
$f_{max}=1$ at $\alpha=0.652$. These and other characteristics of
the $f(\alpha)$ spectrum (such as its width at some value $k
f_{max}$, $0 <k<1$) can be used as tools for classification of
time series with multifractal properties. The consequences of not
fully scaling properties of the time series from panel (c) of Fig.
3 are observed in panel (b) of Fig. 5 where the parabolic form of
the spectrum is slightly deformed.  As such a form is different
from the form for which we can make quantitative conclusions we
can use such spectra only for qualitative conclusions and
classifications. From the point of view of genetics the different
spectra correspond to the different genetic characteristics which
the flies from the first generation obtained from the parents with
different genetic heart defects. Thus only on the basis of the
ECGs we can conclude that the two investigated {\sl Drosophila}
flies are affected in different way by the genetic defects
of their parents. Thus the heartbeat dynamics of
Drosophila is connected to its genetics. This conclusion opens the
way for further intensive research on the relation between (i) the
biomechanical properties of the heart of simple and more complex
animals and humans and (ii) their genetic specifications.
\section{Concluding remarks}
In this paper we demonstrate the possibilities and risks when 
the multifractal detrended fluctuation analysis
is applied to  experimentally obtained time series. We have
started with six time series from which only one was carefully
selected in order to demonstrate monofractal and irregular
behaviour. Then step by step we have demonstrated how the
requirements of the method stop the investigation of different
time series at different points as follows. The requirement for
length was not satisfied by the time series for the acceleration
of the tractor in bouncing regime. The investigation of the time
series for the tractor acceleration in impact regime was stopped
at the point where the requirement for scaling of the fluctuation
function was imposed. At this point we allowed us to make a
relative crude approximation for scaling properties of one of the
time series for the heartbeat activity of {\sl Drosophila}. The
above crude approximation has the consequences in the multifractal
spectra and especially in the $f(\alpha)$ spectra. The
monofractality of time series was clearly illustrated by the
$h(q)$ spectrum of the time series from the random number
generator. Finally the $f(\alpha)$ spectra have shown that the
prescribed form of the spectrum for the multifractal behaviour can
be assigned to only one of the six time series.
\par
In conclusion we warn the researcher to be very careful when he or
she applies the MFDFA to time series from real systems. Only small
number of them are appropriate for such analysis. But when this
analysis can be performed it can supply us with much quantitative
and qualitative information about the relations between (i) the
observed dynamics of the investigated complex system and (ii) the
usually not visible processes,
 responsible for this dynamics.
\section{Acknowledgments}
N. K. V. gratefully acknowledges the support by NFSR of Republic of Bulgaria (contract MM 1201/02).
K. S. thanks to the Japanese Society for Promotion of
Science (JSPS) for the support by the Grant-in Aid for Scientific Research $\# 09660269$.
 E.D.Y. thanks the  EC Marie Curie Fellowship Programm
(contract  QLK5-CT-2000-51155) for the support of her research.

\newpage
\begin{center}
{\sc Figure captions}
\end{center}
\begin{itemize}
\item Figure 1: \\
The time series. Panels (a), (b), (c): time series for the
acceleration of an agricultural tractor.  Horizontal axis : time 
(unit for time is 8 ms for the panels (a) and (b) and 1 ms for the
panel (c)). Vertical axis : accelerations (unit $m/s^{2}$).
Panels (d) and (e): characteristic time series for the
heartbeat activity (ECG)  of {\sl Drosophila melanogaster}. Unit for
time here is 1 ms. Panel (f): time series from the random number
generator {\sf ran2} \cite{nrec}. This generator generates pseudorandom
numbers in the interval [0,1].
\item
Figure 2: \\
Autocorrelation functions for the time series shown in Fig.1.
Panels (a), (b): autocorrelations for the bouncing tractor. Panel (c):
autocorrelations for a tractor working in impact regime. Panels (d), (e):
autocorrelations for the heartbeat activity of {\sl Drosophila
melanogaster}. Panel (f): time series of pseudorandom numbers obtained
by {\sf ran2}. 
\item
Figure3: \\
Fluctuation functions (\ref{nonzero}). For all panels from
bottom to the top the curves (marked with circles) are obtained for
$q=2,4,6,8$. Solid line denote the r.m.s. power law fits of the
corresponding curves. Panel (a): fluctuation functions for the time
series of agricultural tractor in impact regime. Panels (b) and (c):
fluctuation functions for intermaxima time series obtained by the
time series of ECGs of {\sl Drosophila melanogaster}. Panel (d):
fluctuation function for the time series obtained by the generator
{\sf ran2}.
\item
Figure 4: \\
Local Hurst exponent spectra $h(q)$. Panel (a): $h(q)$ spectrum
for the intermaxima time series obtained by the ECG time series of {\sl Drosophila}
shown in panel (d) of Fig.1. Panle (b): $h(q)$ spectrum for the intermaxima
time series obtained by the ECG time series of {\sl Drosophila} shown
in panel (e) of Fig. 1. Panel (c); $h(q)$ spectrum for the time series
obtained by the random number generator {\sf ran2}.
\item
Figure 5: \\
$f(\alpha)$ spectra for the intermaxima time series obtained by the
ECG of {\sl Drosophila melanogaster}. Panel (a): $f(\alpha)$ spectrum for the
intermaxima time series oftained by the ECG of panel (d) of Fig. 1.
Panel (b): $f(\alpha)$ spectrum  for the intermaxima time series obtained
by the ECG of panel (e) of Fig. 1. 
\end{itemize}
\newpage
\begin{figure}[h]
\begin{center}
\includegraphics[angle=270,width=15cm]{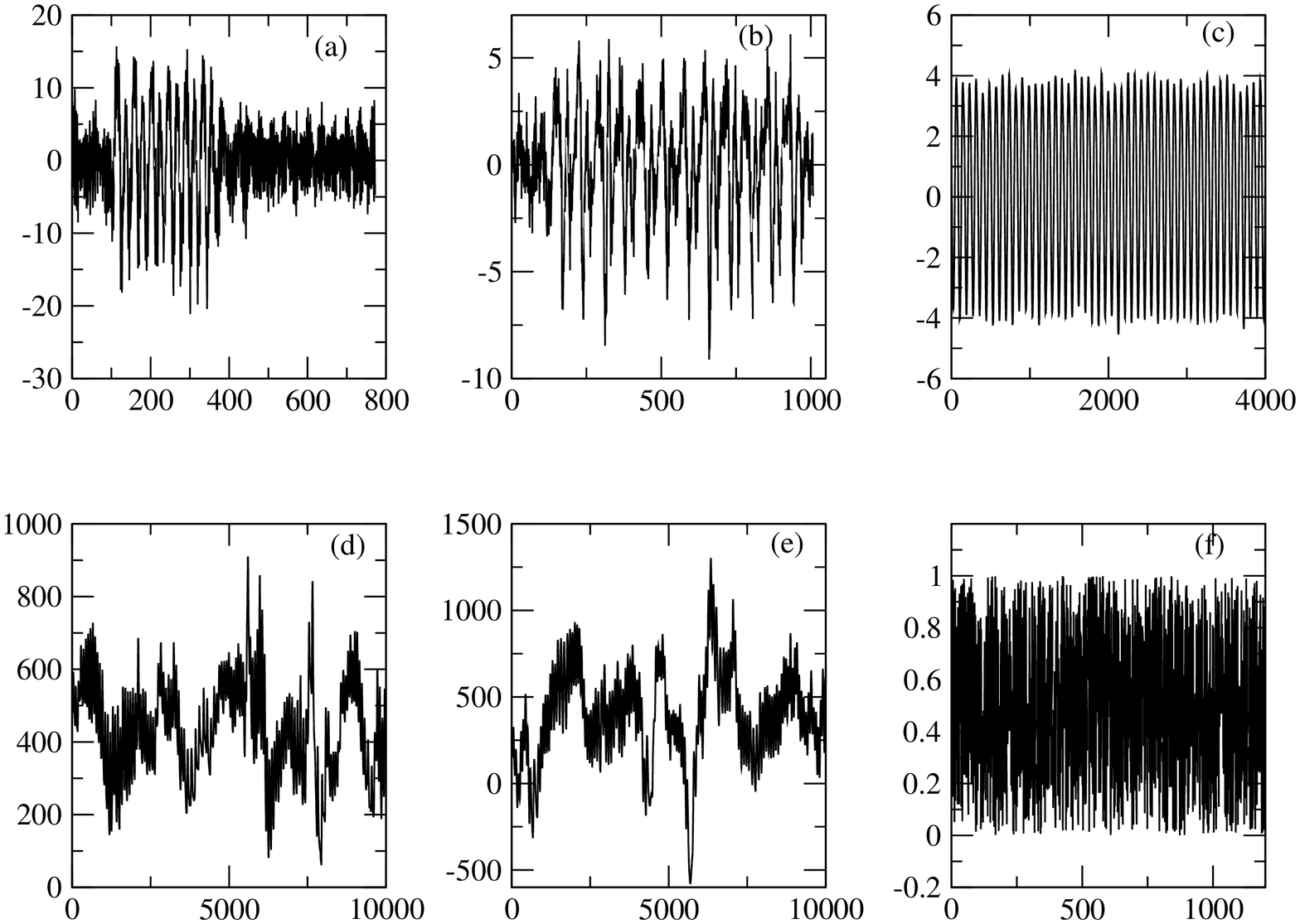}
\end{center}
\end{figure}
\vskip5cm
\begin{center}
Fig. 1
\end{center}
\newpage
\begin{figure}[h]
\begin{center}
\includegraphics[angle=270,width=15cm]{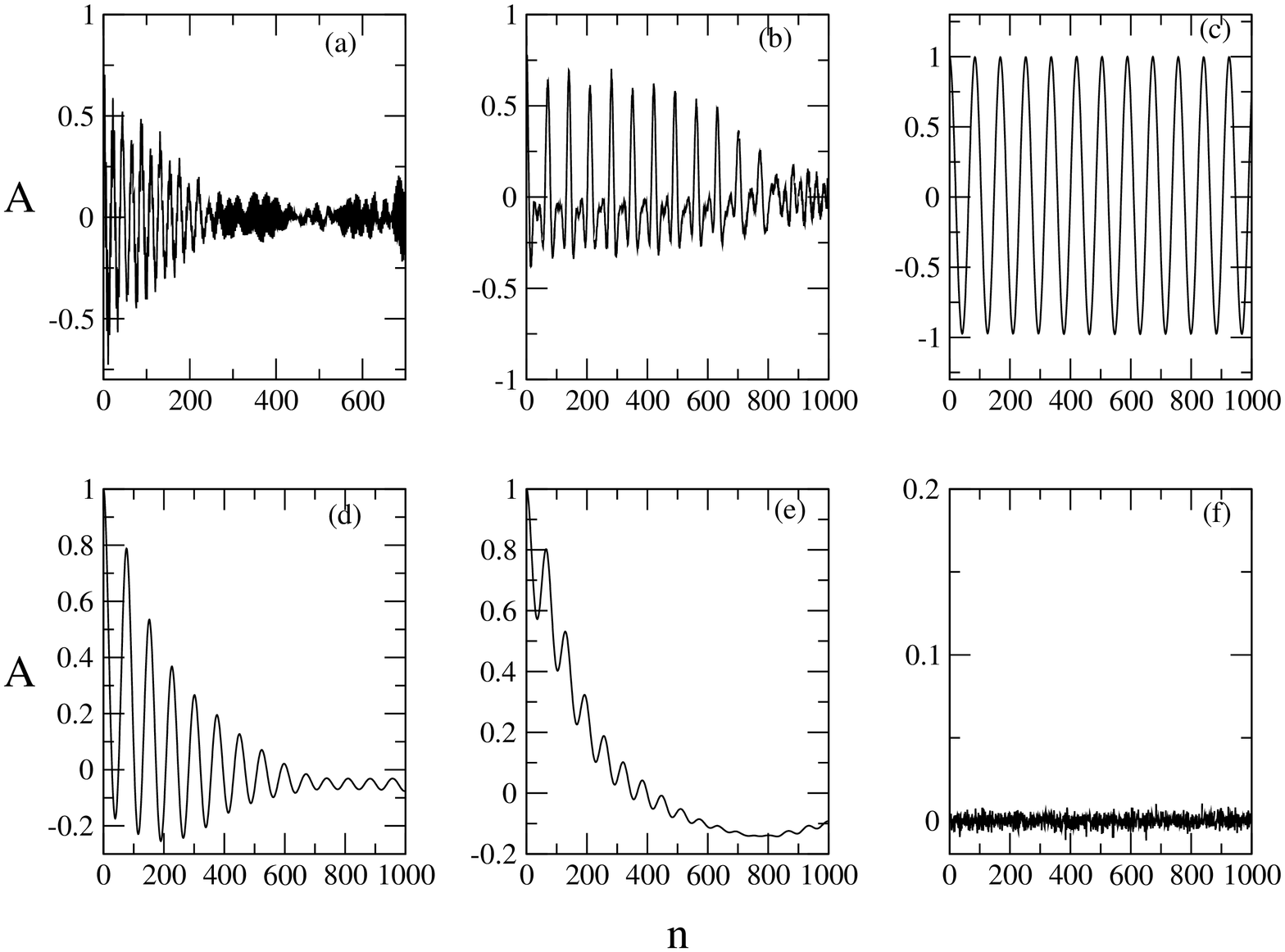}
\end{center}
\end{figure}
\vskip5cm
\begin{center}
Fig. 2
\end{center}
\newpage
\begin{figure}[h]
\begin{center}
\includegraphics[angle=270,width=15cm]{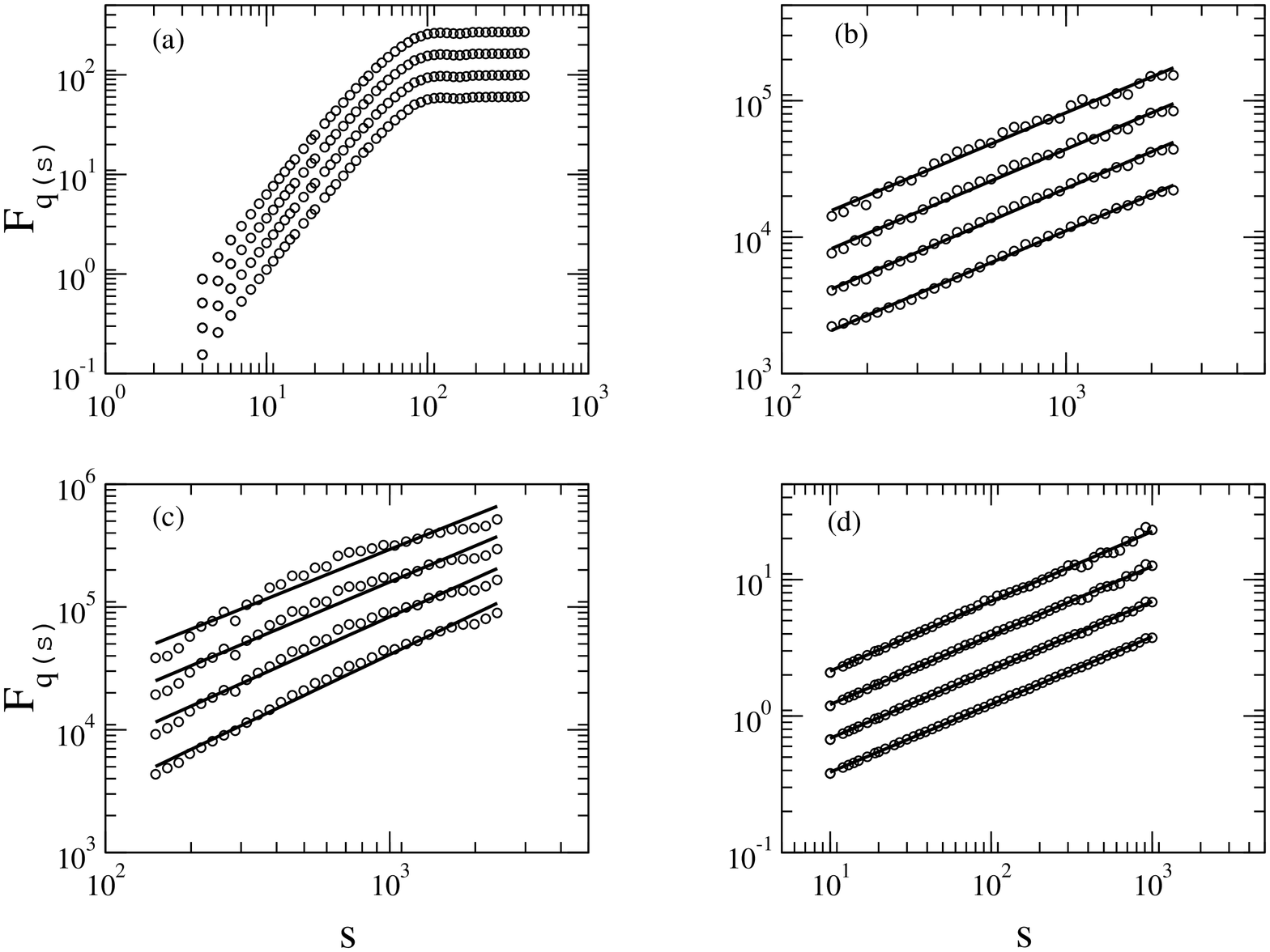}
\end{center}
\end{figure}
\vskip5cm
\begin{center}
Fig. 3
\end{center}
\newpage
\begin{figure}[h]
\begin{center}
\includegraphics[angle=270,width=15cm]{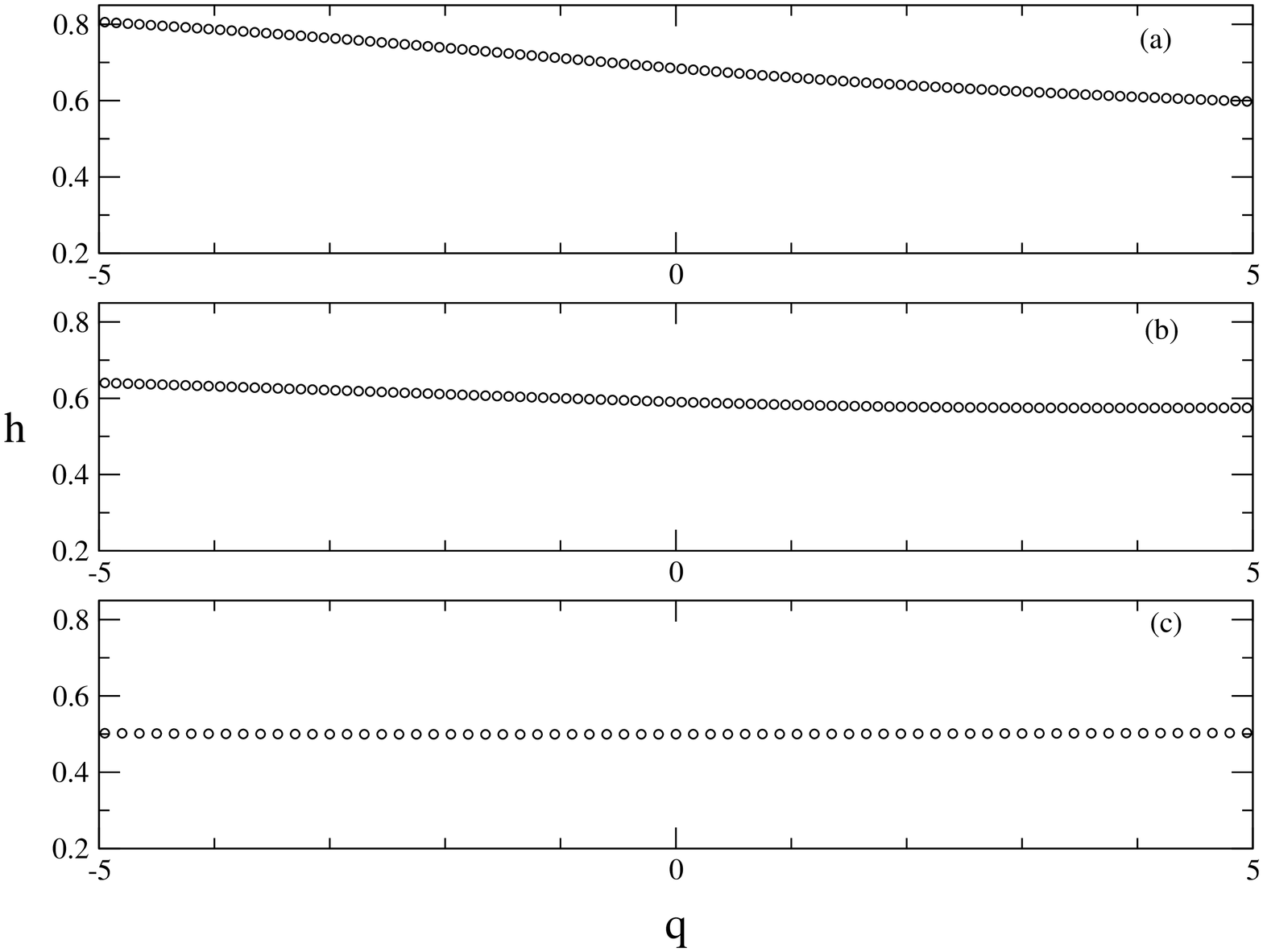}
\end{center}
\end{figure}
\vskip5cm
\begin{center}
Fig. 4
\end{center}
\newpage
\begin{figure}[h]
\begin{center}
\includegraphics[angle=270,width=15cm]{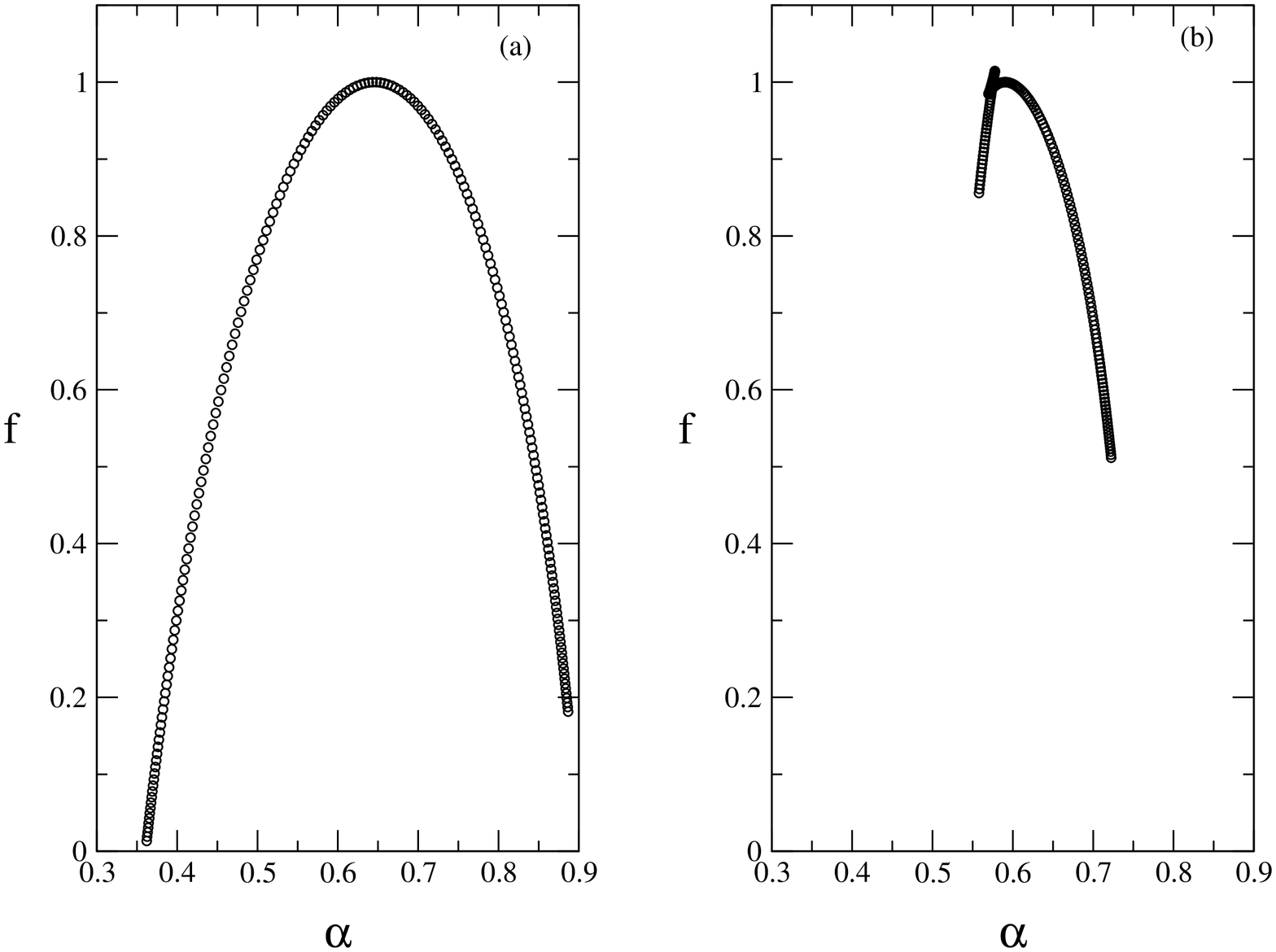}
\end{center}
\end{figure}
\vskip5cm
\begin{center}
Fig. 5
\end{center}

\end{document}